\documentclass[11pt]{article}
\usepackage{graphicx}
\usepackage{epsfig}

\newcommand{\BABARPubYear}    {01}

\newcommand{\BABARProcNumber} {75}
\newcommand{\SLACPubNumber} {9041}
\newcommand{\LANLNumber} {0000}

\newcommand{\babar}{{\sc  $B$a$B$ar~}}
\newcommand{\jpsi}{{$J/\psi~$}}
\newcommand{\psitwo}{{$\psi(2S)~$}}
\newcommand{\chic}{{${\chi_{c}}~$}}
\newcommand{\chicu}{{$\chi_{c1}~$}}
\newcommand{\chicd}{{$\chi_{c2}~$}}
\newcommand{\kpm}{{$K^{\pm}~$}} 
\newcommand{\kz}{{$K_s^{0}~$}} 
\newcommand{\kstpm}{{$K^{*\pm}~$}} 
\newcommand{\kstz}{{$K^{*0}~$}} 
\newcommand{\piz}{{$\pi^{0}~$}} 
\newcommand{\ufours}{{$\Upsilon(4S)~$}} 
\newcommand{\ee}{{$e^+e^-~$}} 
 
\newcommand{\pp}{{$\pi^+\pi^-~$}} 
\newcommand{\pzpz}{{$\pi^0\pi^0~$}} 
\newcommand{\deltae}{{$\Delta E~$}} 
\newcommand{\bibibar}{{$B\overline{B}~$}}

\input pubboard/babarsym

\long\def\inst#1{\par\nobreak\kern 4pt\nobreak
    {\it #1}\par\vskip 10pt plus 3pt minus 3pt}

\begin{document}
{\pagestyle{empty}

\begin{flushright}
SLAC-PUB-\SLACPubNumber \\
\babar-PROC-\BABARPubYear/\BABARProcNumber \\
hep-ex/\LANLNumber \\
December, 2001 \\
\end{flushright}

\par\vskip 4cm

\begin{center}
\Large \bf Studies of B decays to Charmonium at \Lbabar\ 
\end{center}
\bigskip

\begin{center}
\large 
G. Calderini\\
Stanford Linear Accelerator Center \\
2575 Sand Hill rd. Menlo Park, CA 94025 \\
(for the \lbabar\ Collaboration)
\end{center}
\bigskip \bigskip

\begin{center}
\large \bf Abstract
\end{center}
Using 22.7 million $B\overline{B}$ events recorded with the \babar ~detector,
the inclusive branching fractions for the production of $J/\psi$, $\psi(2S)$ 
and $\chi_c$ in $B$ decays are presented. Combining the charmonium state 
with a $K^{\pm}$, $K^0$, $K^{*\pm}$, $K^{*0}$ or $\pi^0$, 
$B$ decays are reconstructed 
exlusively and branching fractions are determined. A preliminary
study is also presented for the $B \rightarrow \eta_c K$ decay mode. 
\vfill
\begin{center}
Contributed to the Proceedings of HEP2001 \\ 
International Europhysics Conference on High Energy Physics, \\
7/12/2001---7/18/2001, Budapest, Hungary
\end{center}

\vspace{1.0cm}
\begin{center}
{\em Stanford Linear Accelerator Center, Stanford University, 
Stanford, CA 94309} \\ \vspace{0.1cm}\hrule\vspace{0.1cm}
Work supported in part by Department of Energy contract DE-AC03-76SF00515.
\end{center}
\section{Introduction}
Reconstruction and study of charmonium mesons in $B$ decays is a crucial component of the measurement 
of time-dependent CP-violating asymmetries\cite{cp}. 

The analyses described in the following paper are based on a sample of 20.7 fb$^{-1}$ collected with \babar ~at the
\ufours resonance with an additional 2.6 fb$^{-1}$ collected below the \bibibar threshold.  
A determination of the $B$ meson branching fractions depends upon an accurate measurement of the number 
of $B$ mesons in the data sample. The number of \bibibar events is determined by comparing the rate of
multi-hadron events in data collected both on and off resonance.
The continuum contribution to the on-resonance sample is estimated by rescaling the number of off-resonance 
hadronic events by the ratio of the number of observed $\mu^+\mu^-$ events in the two samples. 
This procedure yields a total of $22.72 \pm 0.36$ million \bibibar events.

\section{Inclusive decays of $B$ to states containing Charmonium}
\jpsi candidates are selected by requiring two identified leptons of opposite charge. Electrons are 
identified based on the ratio $E/p$ of the energy deposited in the calorimeter to the measured momentum 
from tracking information, the shape of the calorimetric cluster and the ionization in the tracking
detectors. Muons are identified by requiring a minimum ionizing signal in the calorimeter. In addition, the 
shape and penetration of the distribution of hits in the instrumented flux return are used.   
The number of \jpsi events is determined by fitting the invariant mass distribution to a probability density function  
obtained from a simulation including contribution from both final state radiation and bremsstrahlung. 
The fit
yields $15739 \pm 177$ \jpsi $\rightarrow e^+e^-$ and  $13683 \pm 154$ \jpsi $\rightarrow \mu^+\mu^-$ 
signal events (Figure \ref{fig1}).
\begin{figure}[h]
\begin{center}
\begin{tabular}{cc}
\epsfig{file=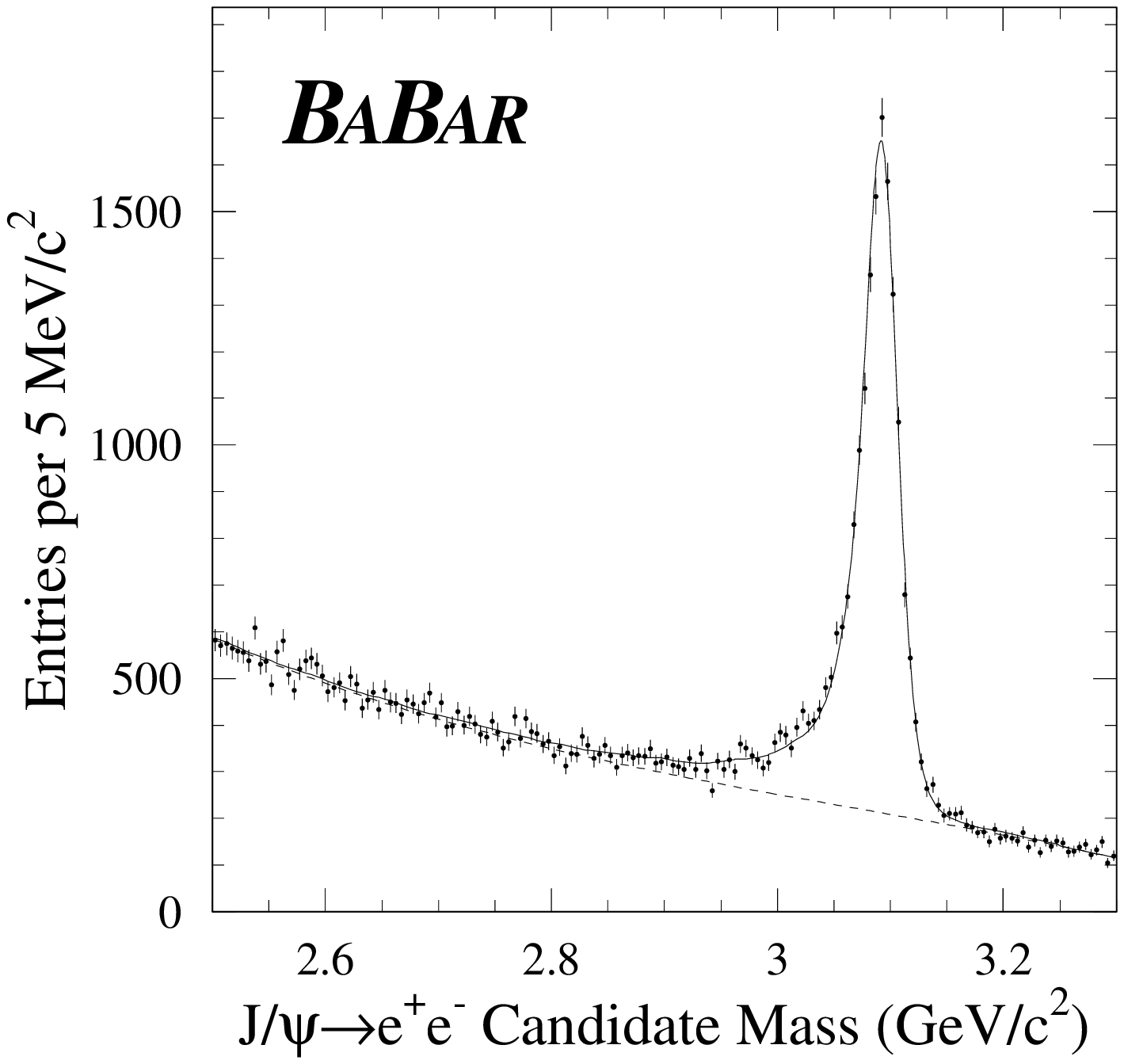,width=6.0cm} &
\epsfig{file=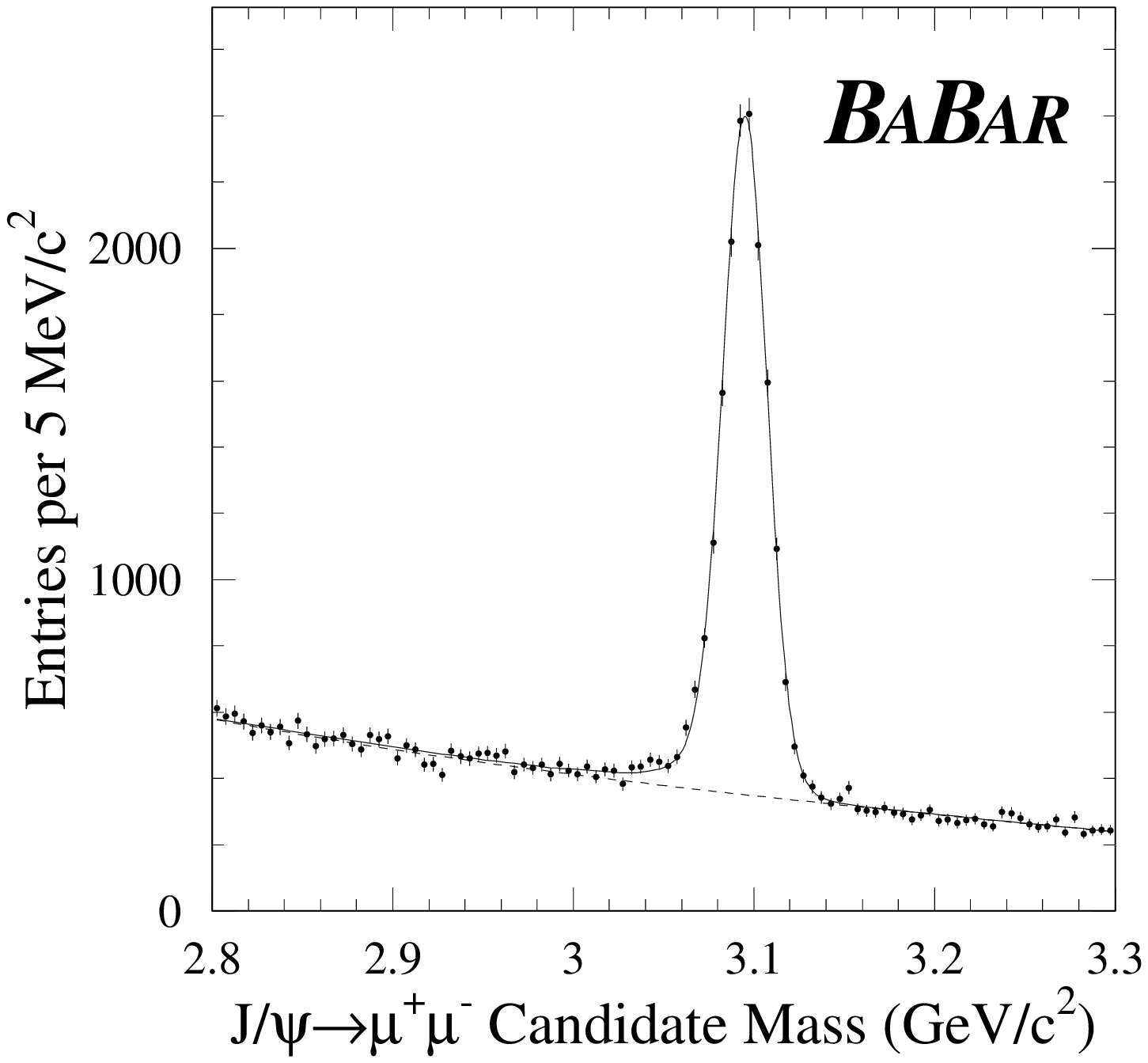,width=6.0cm} 
\end{tabular}
\end{center}
\caption{Invariant mass distributions for inclusive $J/\psi$ production in $B$ decays}
\label{fig1}         
\end{figure}

The \psitwo candidates are reconstructed in the \psitwo $\rightarrow l^+l^-$ 
and \psitwo $\rightarrow J/\psi\pi^+\pi^-$ decays. In the latter case the signal 
yield is extracted by a fit 
to the mass difference between the \psitwo and the \jpsi reconstructed candidates. We find 
$552 \pm 50$ decays to \ee, $437 \pm 44$ decays to $\mu^+\mu^-$, $405 \pm 37$ decays to 
$J/\psi(e^+e^-)\pi^+\pi^-$ and $400 \pm 34$ decays to  $J/\psi(\mu^+\mu^-)\pi^+\pi^-$.        

The \chicu and \chicd candidates are selected by combining a reconstructed \jpsi with a photon.
The signal yield is determined by fitting the mass difference between the \chic ~and the
\jpsi candidates. The shape of the signal is extracted from Monte Carlo, with the mass difference 
between the \chicu and \chicd peaks fixed to the PDG value \cite{pdg}. 
The fit gives $476 \pm 71$ \chicu and $86 \pm 59$ \chicd candidates for the $J/\psi \rightarrow e^+e^-$
decay and  
$545 \pm 60$ \chicu and $104 \pm 56$ \chicd candidates for the $J/\psi \rightarrow \mu^+\mu^-$ decay.

\begin{table}[h]
\begin{center}
\begin{tabular}{||c|c||}
\hline
Mode & Br($\times 10^{-2}$) \\
\hline
$B\rightarrow J/\psi~ X$ & $1.044 \pm 0.013 \pm 0.028$ \\
$B\rightarrow \psi(2s)~ X$  & $0.275 \pm 0.020 \pm 0.029$ \\ 
$B\rightarrow \chi_{c1} X$  & $0.378 \pm 0.034 \pm 0.026$ \\
$B\rightarrow \chi_{c2} X$  & $<0.21~(90\%~CL)$  \\
\hline
\end{tabular}
\end{center}
\caption{Measured branching fractions for inclusive charmonium production in $B$ decays.}
\label{incl_tab}
\end{table}
Values for the branching fractions are extracted from the yields\cite{inclu_paper}. 
A $90\%$ CL limit is set on the $B$ decay to \chicd.
The results for the inclusive production of \jpsi, \psitwo and \chic ~in
$B$ decays are summarized in Table \ref{incl_tab}, where the first uncertainty is
statistical error and the second is systematics.

\section{Exclusive decays of $B$ to Charmonium}
The reconstruction of exclusive decay modes containing charmonium presents 
a very low background in most of the channels. For this reason the lepton 
identification criteria are loosened
for one of the two \jpsi decay products. As in the inclusive analysis, \psitwo candidates
are reconstructed by their decay to \ee, $\mu^+\mu^-$ and $J/\psi~\pi^+\pi^-$. The \chic ~candidates
are selected through their decay to $J/\psi \gamma$. 

The charmonium states are selected in a window around their expected mass\cite{pdg} 
for the decays to leptons. In the decays to \jpsi states the mass difference 
distribution between the charmonium candidate and the reconstructed \jpsi is used 
instead. 
  
Selected candidates are then paired with a $K^+$, \kz (either \pp or \pzpz), 
$K^{*+}$ (either $K^+$\piz or \kz$\pi^+$), \kstz (either $K^+\pi^-$ 
or  \kz\piz), \piz or $K_L$ to form a $B$ candidate. The two most significant observables used to 
identify the signal are the difference \deltae in the center-of-mass frame between the 
reconstructed $B$ energy and half the nominally available energy, $\sqrt{s}/2$, and the energy-substituted mass, 
$m_{ES}= \sqrt{s/4-{P_B^{*2}}}$, where $P_B^{*}$ is the center-of-mass momentum of the
$B$ candidate. A sample of these distributions is given in Figure \ref{fig_de_mes} for $J/\psi K_S$ events.

\begin{figure}
\begin{center}
\begin{tabular}{cc}
\epsfig{file=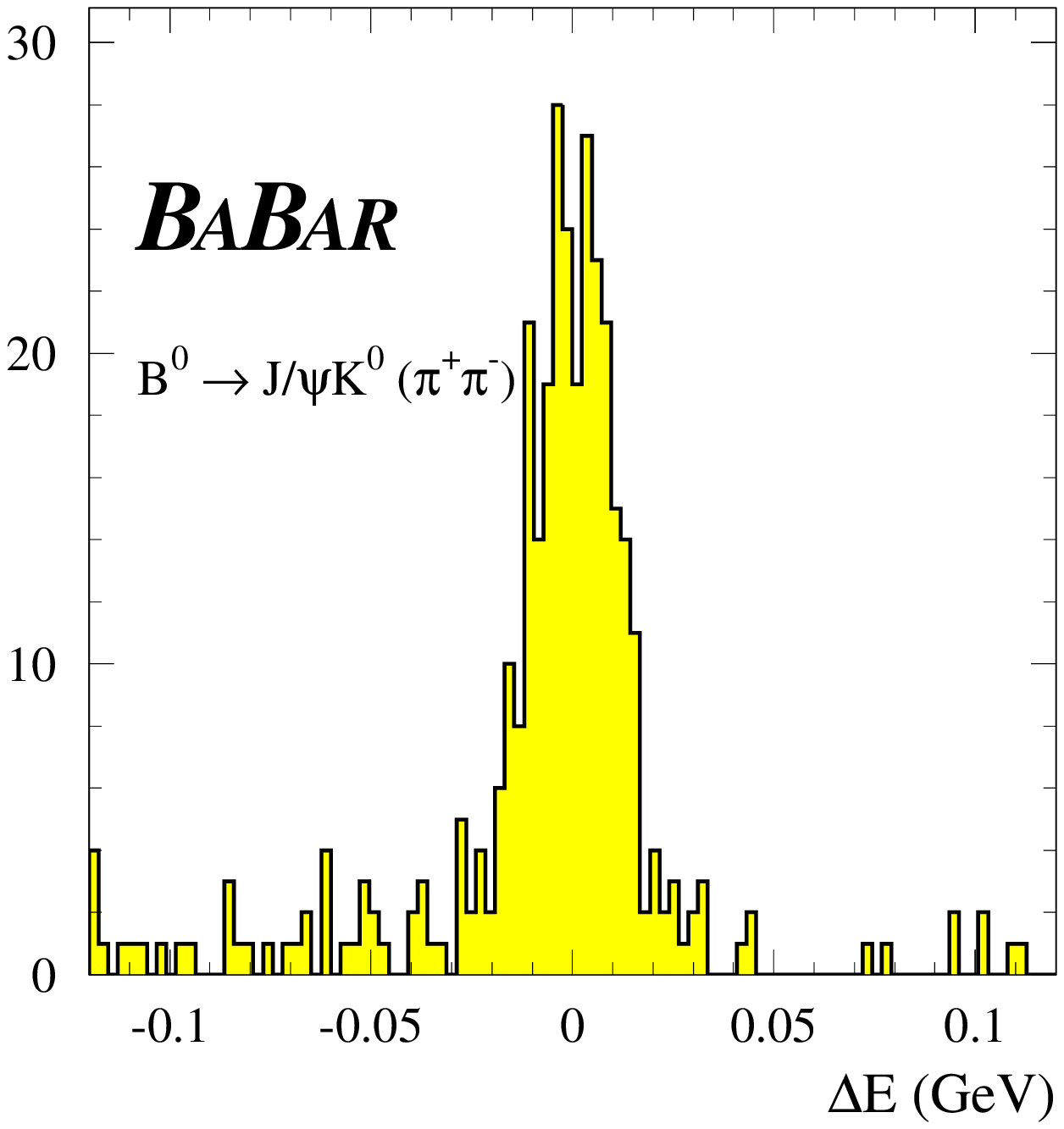,width=6.0cm} &
\epsfig{file=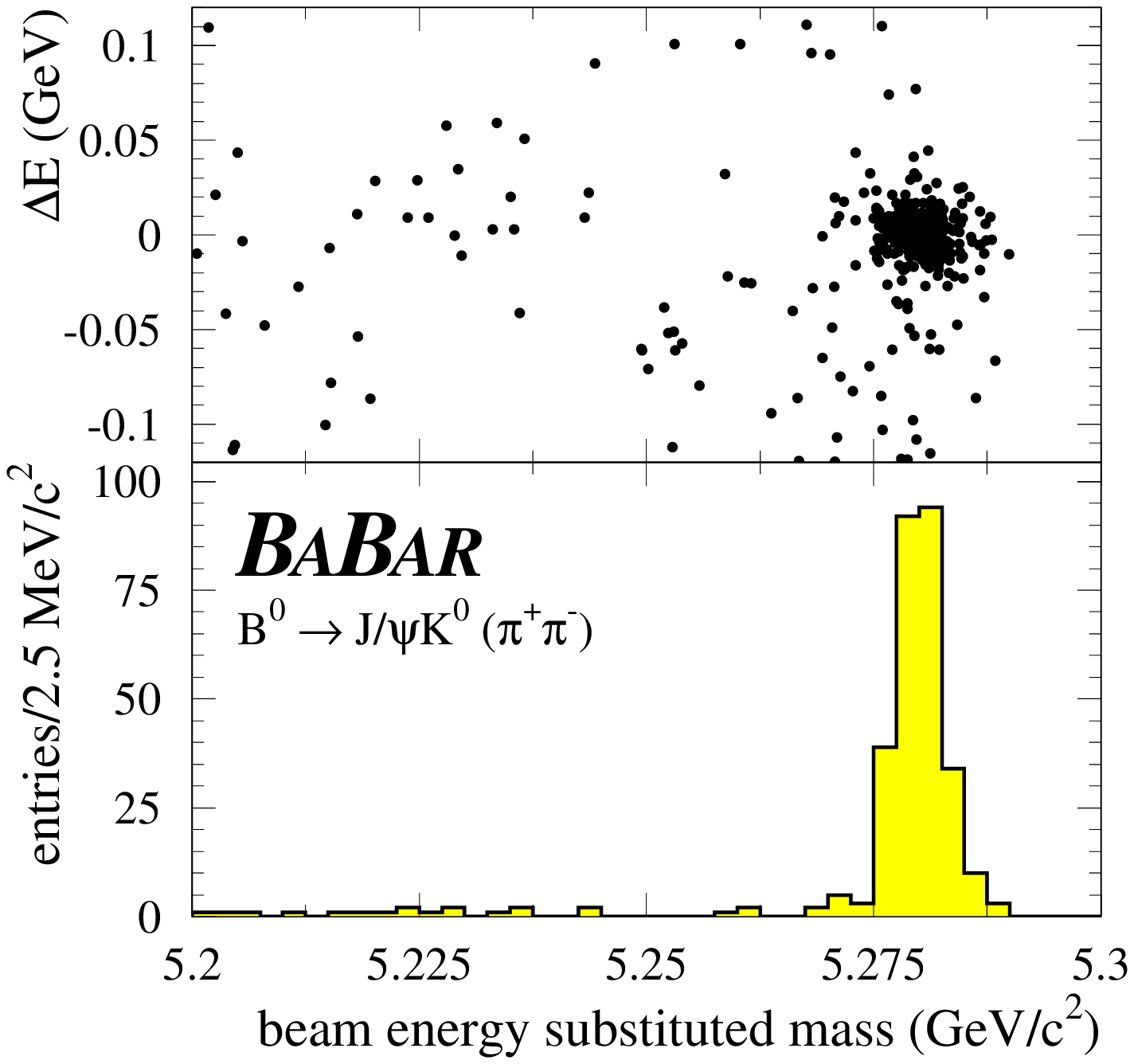,width=6.0cm} 
\end{tabular}
\end{center}
\caption{Example of the \deltae (left) and $m_{ES}$ (right) distributions for the decay
$B^0 \rightarrow J/\psi K_s^0$ }
\label{fig_de_mes}         
\end{figure}

In the case of multiple candidates per event, only the candidate with the smallest
$|\Delta E|$ is selected. 
For all modes except $B \rightarrow J/\psi K_L$ and $B \rightarrow J/\psi K^*$, the number
of signal events is determined from the observed number of events in the $(\Delta E, m_{ES})$
region after background subtraction. In addition to the usual combinatorial component, the 
background distribution shows an excess of events in the signal region. 
The combinatorial contribution is estimated by using an ARGUS function 
in the fit to the $m_{ES}$ distribution. The peaking background component 
is obtained from simulation of inclusive $B$ decays to charmonium, 
after removing the signal events.

The signal yields for the $K^{*0}$ and $K^{*+}$ modes are determined simultaneously from a 
likelihood fit, which is needed to account for the cross-feed between the $K^*$ decay
channels.  

A different technique is used for the $B\rightarrow J/\psi K_L$ decay mode.
In this case only the $K_L$ direction is measured with information from the calorimeter
and the instrumented flux return. Given this direction and the reconstructed charmonium 
candidate, the $K_L$ energy is extracted by using the $B$ mass as a constraint.  
To eliminate cross-feed from other decay modes, a veto has been introduced for events
which have been selected already in other exclusive modes. This procedure yields a
purity of about $50\%$. Due to this method, the $m_{ES}$ distribution cannot 
be used to determine the signal yield. The $\Delta E_{J/\psi K^0_L}$ distribution is 
used in a log-likelihood fit. The shapes of the signal and inclusive charmonium 
background components are taken from Monte Carlo simulations. The shape of the non-charmonium 
background component is taken from an ARGUS fit to the $\Delta E_{K^0_L}$ distribution for 
events in the \jpsi mass sideband. After the background subtraction, this channel gives
a signal yield of $183 \pm 14$ events (Figure \ref{fig_etac}, left plot).
\begin{figure}[bht]
\begin{center}
\begin{tabular}{cc}
\epsfig{file=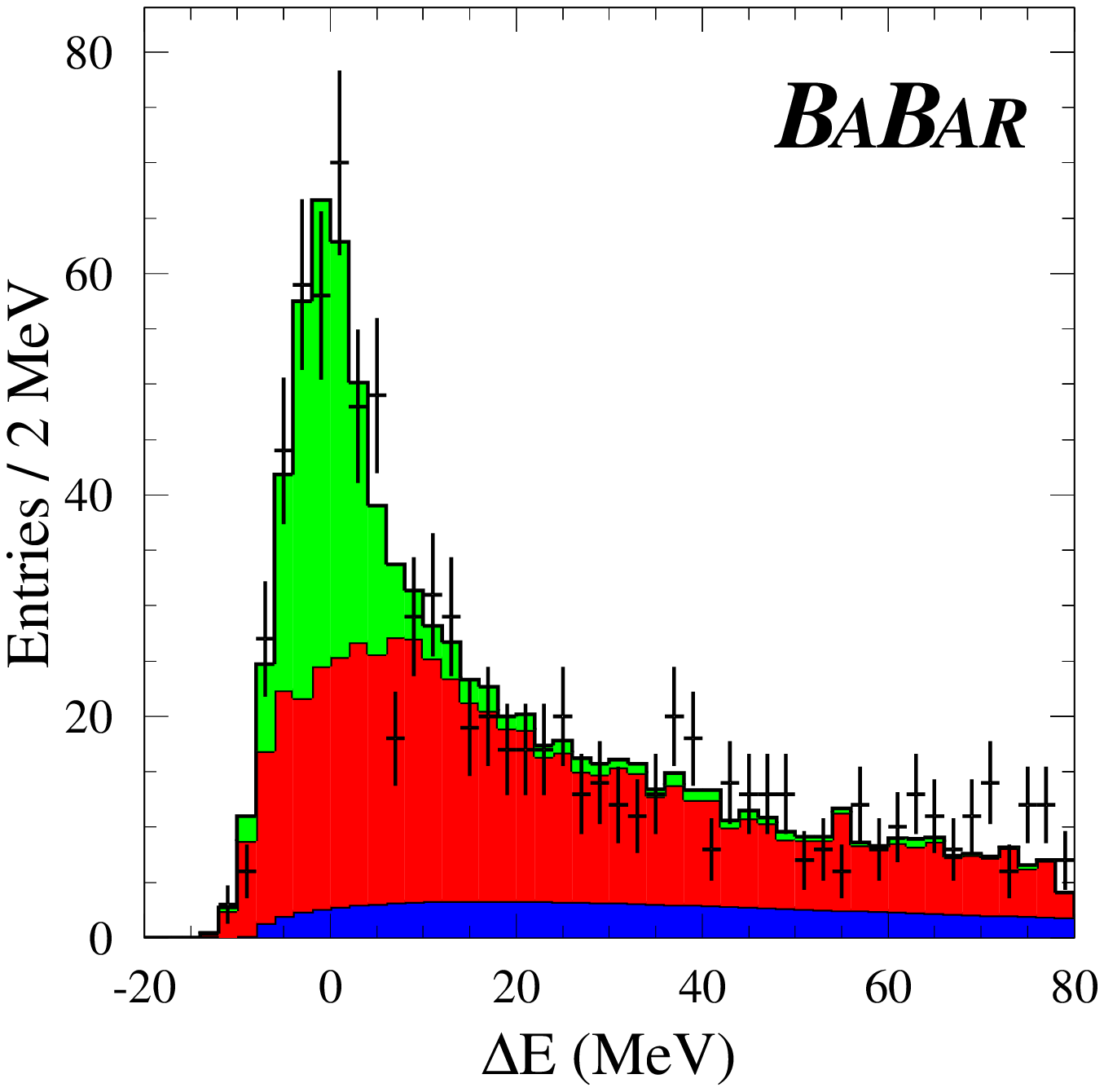,width=6.3cm} &
\epsfig{file=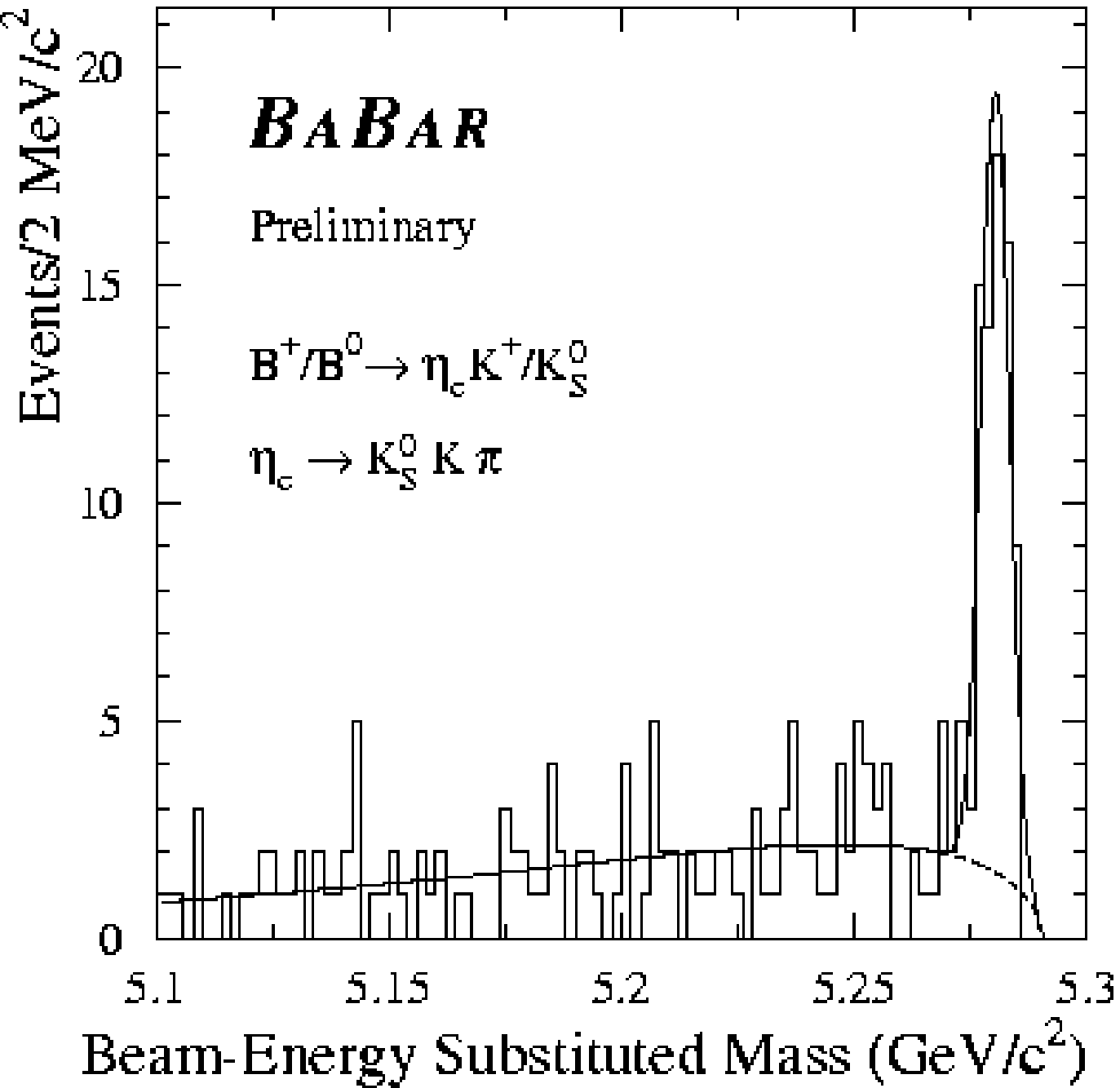,width=6.3cm}  
\end{tabular}
\end{center}
\caption{(Left plot) \deltae distribution for $B^0 \rightarrow J/\psi K_L$ decays. Points are data,
solid line is the Monte Carlo simulation. The three components are respectively
signal, background events which include a real $J/\psi$, and non-$J/\psi$ background.
(Right plot) preliminary study of $B$ decay to $\eta_c K$. In the above plot the $\eta_c$ candidate
is reconstructed through $K_S K \pi$ decays.}
\label{fig_etac}
\end{figure}
\begin{table}[htb]
\begin{footnotesize}
\begin{center}
\begin{tabular}{||ll|c|c||}
\hline
Mode & & Yield & Br($\times 10^{-4}$) \\
\hline
$B^0 \rightarrow J/\psi K^0$       & $K_S^0\rightarrow\pi^+\pi^-$ & $265.5 \pm 2.9$ & $8.5  \pm 0.5  \pm 0.6$  \\
                                   & $K_S^0\rightarrow\pi^0\pi^0$ & $62.5  \pm 3.8$ & $9.6  \pm 1.5  \pm 0.7$  \\
                                   & $K_L^0$                      &  $183 \pm 14$   & $6.8  \pm 0.8  \pm 0.8$  \\
                                   & All                          &                 & $8.3  \pm 0.4  \pm 0.5$  \\
$B^+ \rightarrow J/\psi K^+$       &                              &  $1109 \pm 4 $  & $10.1 \pm 0.3  \pm 0.5$  \\
$B^0 \rightarrow J/\psi \pi^0$     &                              &  $13.6 \pm 0.9$ & $0.20 \pm 0.06 \pm 0.02$ \\
$B^0 \rightarrow J/\psi K^{*0}$    &                              &  $594 \pm 8.5$  & $12.4 \pm 0.5  \pm 0.9$  \\
$B^+ \rightarrow J/\psi K^{*+}$    &                              & $377.4 \pm 16.9$ & $13.7 \pm 0.9  \pm 1.1$  \\
$B^0 \rightarrow \psi(2S) K^{0}$   &                              &  $56.0 \pm 3.4$ & $6.8  \pm 1.0  \pm 1.1$  \\
$B^+ \rightarrow \psi(2S) K^{+}$   &                              & $207.3 \pm 6.2$ & $6.3  \pm 0.5  \pm 0.8$  \\
$B^0 \rightarrow \chi_{c1}K^{0}$   &                              & $26.1  \pm 2.5$ & $5.4  \pm 1.4  \pm 1.1$  \\
$B^+ \rightarrow \chi_{c1}K^{+}$   &                              & $145.1 \pm 7.2$ & $7.5  \pm 0.8  \pm 0.8$  \\
$B^0 \rightarrow \chi_{c1}K^{*0}$  &                              & $32.6  \pm 6.0$ & $4.8  \pm 1.4  \pm 0.9$  \\
$B^0 \rightarrow J/\psi\pi^+\pi^-$ &                              & $29.1 \pm 9.4 $ & $0.46 \pm 0.11 \pm 0.08$ \\
\hline
\end{tabular}
\end{center}
\end{footnotesize}
\caption{Summary of the signal yields and extracted branching fractions for the different $B$ decays reconstructed 
into exclusive charmonium final states by the \babar ~analyses described in this paper.}
\label{br_table}
\end{table}

The information on the signal yields and the measured branching fractions for all 
presented exclusive modes\cite{exclu_paper} is summarized in Table \ref{br_table}, 
while Figure \ref{pdgco} shows 
a comparison of the new preliminary \babar ~results to the current PDG values. 
\begin{figure}[h!]
\begin{center}
\epsfig{file=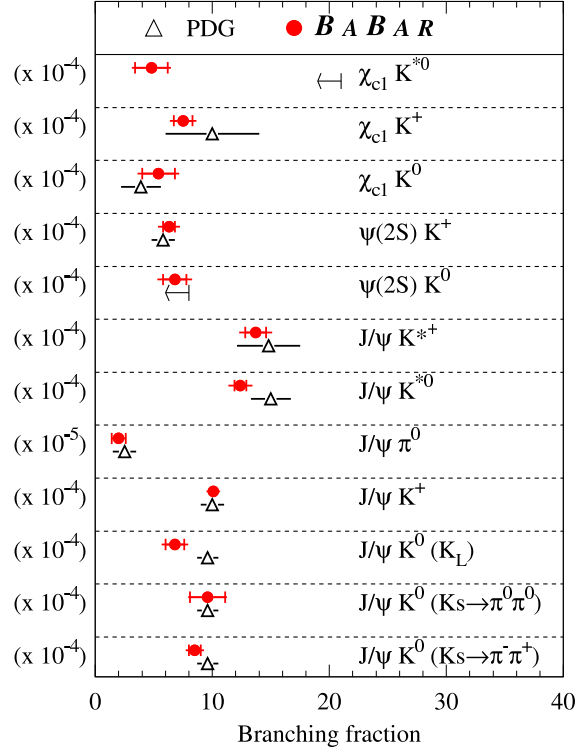,height=10.5cm}
\end{center}
\caption{Comparison of the preliminary results from the \babar ~summarized in the previous table and the 
published PDG values.}
\label{pdgco}
\end{figure}

In addition to the analyses described above, a very preliminary study has been recently
performed by \babar, on exclusive reconstruction of $B$ decays to $\eta_c K$ modes.
The $\eta_c$ has been studied in the $K_S K^{+} \pi^{-}$, $K^+ K^- \pi^{0}$ and
$K^+ K^- K^+ K^-$ ($\phi\phi$ plus non-resonant) decay modes. The energy-substituted mass
distribution for the decay $\eta_c \rightarrow K_S K^{+} \pi^{-}$ is presented as an example
in Figure \ref{fig_etac}, 
right plot. No value has been extracted yet for the branching fractions
in these decay modes.

\section{Summary}
Using 22.7 million \bibibar events recorded by the \babar ~detector, the inclusive branching fractions for
the production of \jpsi, \psitwo and \chic ~are presented. Combining the charmonium state with either 
a \kpm, $K^0$, \kstpm, \kstz or $\pi^0$, $B$ decays are reconstructed exclusively and their branching fractions 
are determined.



\end{document}